# Holotomography in 2025: From Morphometric Imaging to AI-Driven Multimodal Phenotyping

YongKeun Park[1,2,3,*]

[1]Department of Physics, KAIST, Daejeon, Republic of Korea
[2]KAIST Institute for Health Science and Technology, KAIST, Daejeon, Republic of Korea
[3]Tomocube Inc., Daejeon, Republic of Korea
E-mail: yk.park@kaist.ac.kr

**Abstract**

By 2025, holotomography (HT) has matured from a niche optical modality into a versatile platform for quantitative, label-free imaging in biomedicine. By reconstructing the three-dimensional refractive-index (RI) distribution of cells and tissues, HT enables high-resolution volumetric imaging with low phototoxicity and minimal sample perturbation. This Review surveys recent advances in the field and highlights three emerging directions: (i) the incorporation of deep-learning approaches for virtual staining, phenotypic classification, and automated analysis; (ii) the extension of HT to structurally complex biological systems, including organoids and thick tissue specimens; and (iii) the integration of HT with complementary modalities, such as Raman and polarization-sensitive microscopy, to enhance molecular and biophysical specificity. We summarize current HT applications spanning subcellular phenotyping, metabolic and mechanical profiling, and early-stage clinical studies in areas such as infectious disease and pathology. Finally, we discuss remaining technical and translational challenges and outline a roadmap for the prospective integration of HT into digital pathology and high-throughput screening workflows.

## 1. Introduction

Quantitative Phase Imaging (QPI) has long promised to reveal the intrinsic physical contrast of biological specimens without the need for exogenous labels, offering a fundamentally non-perturbative window into cellular structure and composition [1]. In 2025, holotomography (HT)—a three-dimensional implementation of QPI [2]—has largely fulfilled this promise by overcoming long-standing limitations in acquisition speed, specificity, and data interpretability [3,4]. Through volumetric reconstruction of refractive index (RI) distributions, HT now enables quantitative, label-free imaging of living systems with sufficient resolution, throughput, and robustness to address complex biological questions.

HT has expanded far beyond its early applications in isolated or weakly scattering circulating cells or cell cultures [5,6]. Recent studies demonstrate its ability to interrogate biological organization across scales, from whole-organism microanatomy in tardigrades [7] to metabolic and structural heterogeneity in patient-derived cancer organoids [3,8]. These advances reflect a broader conceptual shift: HT is no longer viewed merely as a contrast-enhanced imaging modality, but as a physical phenotyping platform capable of capturing density, mass distribution, and spatial organization in three dimensions.

This transformation has been accelerated by the convergence of HT with artificial intelligence (AI) [9]. Deep learning frameworks trained on RI tomograms have enabled "organelle-aware" representation learning, allowing molecularly meaningful phenotypes to be inferred from purely physical measurements [10,11]. Such approaches bridge the historical gap between label-free imaging and molecular specificity [12–15], redefining how structural and functional information can be extracted from QPI data.

The year 2025 also marked a pronounced diversification of HT hardware and reconstruction strategies. Novel optical configurations, exemplified by Multi-Pattern Sparse Axial-Scanning (MuPaSA), have broken the traditional speed–accuracy trade-off, enabling high-fidelity volumetric imaging of rapid cellular dynamics at subcellular resolution [16]. In parallel, polarization-sensitive HT has introduced intrinsic specificity to birefringent structures such as lipid droplets [17], while low-coherence and multimodal implementations have extended HT to thick, scattering samples and correlative morpho-chemical analysis [3,18].

These technological advances have been matched by a surge in biological and preclinical applications. In 2025, HT has been routinely employed to quantify antimicrobial efficacy against multidrug-resistant bacteria [19], assess the quality of cryopreserved reproductive cells [20], and visualize infection dynamics in complex organoid-based models of zoonotic viruses [21]. Collectively, these studies highlight the growing role of HT as a generalizable, species-agnostic imaging platform for dynamic and heterogeneous biological systems.

In this Review, we survey 64 HT–related studies reported in 2025, encompassing both methodological developments and biological or preclinical applications (Table 1 and Fig. 1). The literature is organized along two orthogonal axes: application domain–cell biology and biophysics; organoids, stem cells, and regenerative medicine; clinical and cancer biology; and microbiology and infection—and analytical strategy, including quantitative phenotyping, three-dimensional and live or structurally complex analysis, and AI–based approaches. Reference numbers correspond to the bibliography, and studies spanning multiple domains or analytical strategies are included in all relevant categories. On the basis of this structured framework, we synthesize recent advances across three interrelated dimensions—hardware innovation, AI-enabled analysis, and emerging application spaces—and conclude by outlining a strategic roadmap for the continued evolution of HT toward standardized, interpretable, and clinically actionable physical phenotyping.

|  | *1. Quantitative Phenotyping* | *2. 3D / Live / Complex* | *3. Artificial Intelligence* |
|---|---|---|---|
| *A. Cell biology and biophysics* | [4,18,20,22–24,24–31] | [10,32,33] | [34] |
| *B. Organoids, and stem cells* | [35,36] | [7,8,11,16,21,26,37–43] | [44] |
| *C. Clinical and cancer biology* | [3,17,24,33,45,45–56] | [57,58] | [59,60] |
| *D. Microbiology and infection* | [61–66] | [19,67] | [68] |

**Table 1** | Landscape of HT-based studies published in 2025, categorized by application domain and analytical strategy.

## 2. Technological Frontiers: Speed, Physics, and Multimodality

### 2.1. Breaking the Speed-Accuracy Trade-off

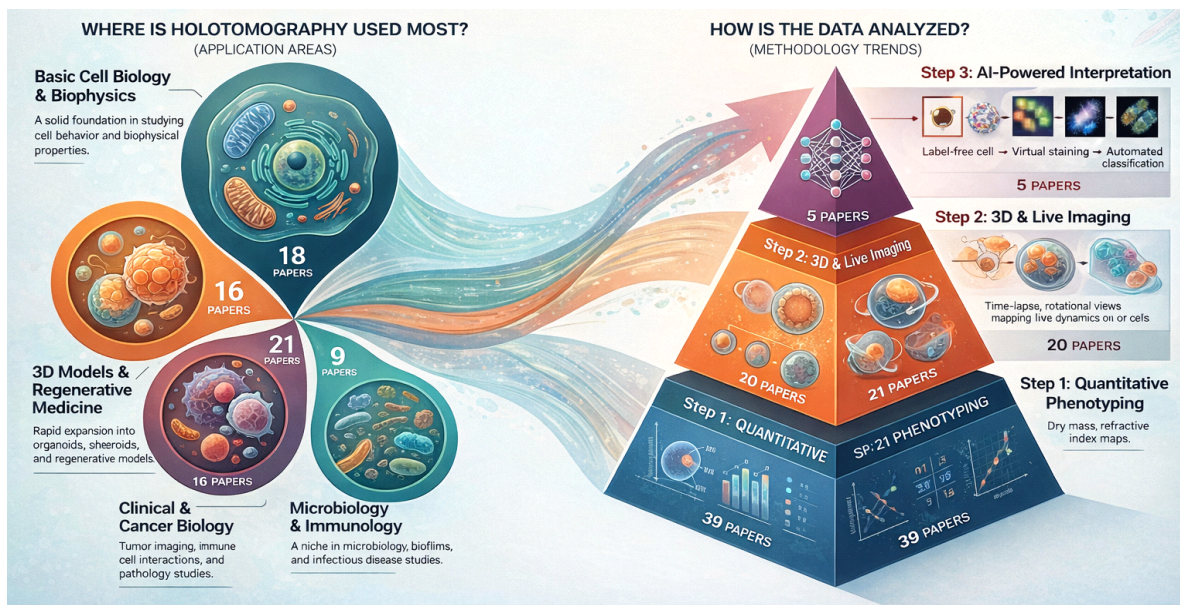

**Figure 1** | **2025 HT trends: the future of label-free imaging.** HT is an advanced microscopy technique that provides label-free, quantitative 3D imaging of live cells and tissues by measuring their RI

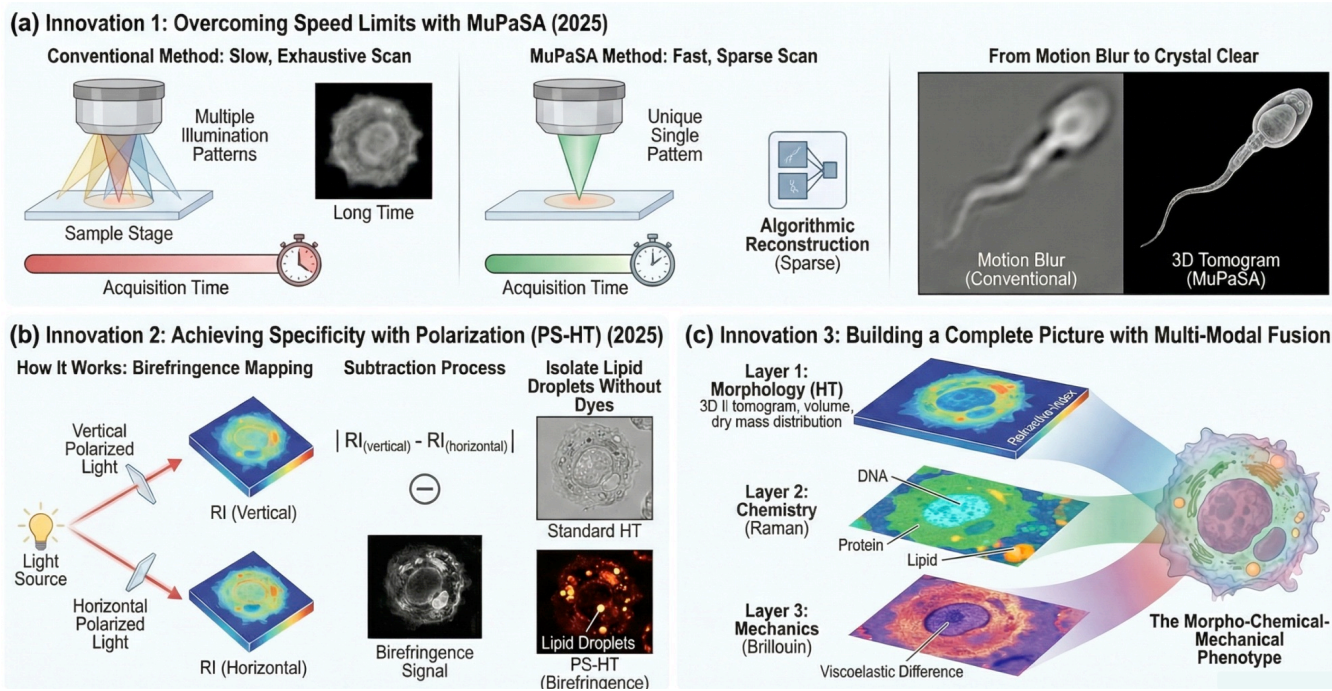

**Figure 2 | Technological frontiers in HT: speed, specificity, and multimodality.** (a) Accelerated volumetric imaging with MuPaSA. By using a single illumination pattern per axial plane and cycling patterns across the z-stack, MuPaSA (Hong et al., 2025) increases volumetric acquisition rates by up to fourfold, enabling 3D imaging of fast cellular dynamics. (b) Label-free specificity via PS-HT. PS-HT (Khadem et al., 2025) exploits lipid-droplet birefringence by acquiring RI tomograms under orthogonal polarization states, producing high-contrast birefringence maps for robust, threshold-independent lipid segmentation without labels. (c) Multimodal morpho-chemical phenotyping. Integration of HT with Raman and Brillouin modalities (Matsushita et al., 2025; Bianchi et al., 2025) enables co-registered 3D morphology, molecular composition, and viscoelastic properties, revealing links between lipid accumulation and chromatin organization. Images are schematic and illustrative; see cited studies for experimental da

A fundamental limitation of conventional HT has been the inherent trade-off between temporal resolution and reconstruction fidelity [69,70] (Fig. 2a). High-quality 3D reconstruction typically requires multiple illumination patterns per axial plane, constraining volumetric imaging speed and limiting the study of rapid biological dynamics. In 2025, the introduction of MuPaSA directly addressed this bottleneck [16]. By illuminating a single, unique pattern at each axial plane while cycling illumination patterns across the z-stack, MuPaSA preserves spatial frequency coverage while reducing acquisition redundancy. As a result, volumetric acquisition rates were increased by approximately fourfold without compromising reconstruction quality.

This advance enabled artifact-free 3D imaging of fast-moving and highly dynamic biological systems, including swimming spermatozoa and rapid cellular responses to osmotic perturbations, which are typically blurred or distorted in conventional HT reconstructions. Beyond acquisition strategies, reconstruction methodologies have also evolved. Generalized Reciprocal Diffractive Imaging (RDI) emerged as a powerful framework for non-interferometric, single-shot quantitative phase imaging [4]. By modulating the Fourier spectrum to suppress the dominant DC component, RDI enables stable, high-fidelity phase reconstruction from intensity-only measurements [71]. This approach offers a robust, reference-free alternative to interferometric HT, particularly well-suited for diffusive samples such as live cells and tissue sections.

### 2.2. Polarization and Birefringence

While RI tomography provides quantitative information on mass density and intracellular organization, RI contrast alone lacks intrinsic chemical specificity. To overcome this limitation, polarization-sensitive HT (PS-HT) was developed to exploit the birefringent properties of anisotropic cellular structures [17] (Fig. 2b). PS-HT measures polarization-dependent phase retardation by acquiring 3D RI tomograms under orthogonal polarization states, enabling direct mapping of birefringence within living cells.

This approach has proven particularly effective for the label-free identification and segmentation of lipid droplets (LDs). Due to the anisotropic arrangement of phospholipid monolayers and cholesteryl ester cores, LDs exhibit characteristic Maltese-cross patterns under polarized illumination. PS-HT leverages this intrinsic optical signature to distinguish LDs from other high-RI organelles that are otherwise indistinguishable in conventional HT. In prostate cancer cells, PS-HT enabled precise, threshold-independent quantification of lipid volume and dry mass, revealing metabolic stress–induced lipid remodeling with high specificity and reproducibility.

The concept of PS-HT can be further extended to measure the 3D distribution of dielectric tensor [72–79]. The technology can be further expanded to biological and preclinical investigation.

*2.3. Low-Coherence and Multimodal Integration*

The synergy between HT and other optical modalities is defining the next generation of bioimaging. Low-coherence HT (LC-HT) has become the standard for imaging thick, scattering samples like organoids, as it significantly reduces speckle noise compared to coherent systems [44] (Fig. 2c). This capability was essential for the long-term, label-free monitoring of intestinal organoids, allowing for the quantification of protein density and swelling without phototoxicity. Furthermore, multimodal approaches combining HT with Raman spectroscopy have enabled the simultaneous acquisition of morphological and biochemical data, facilitating the phenotyping of breast cancer cells [3] and the study of chromatin condensation involving lipids [18]. A study has also combined HT with Brillouin microscopy to correlate lipid accumulation with heterochromatin condensation and viscoelasticity in living cells [32]. Correlative imaging remains essential for validation. It has been used to track mitochondrial dysfunction in neurons [11], monitor nanoparticle uptake [41], and validate specific markers in various cancer models [3].

Beyond 3D structural imaging, the functional versatility of HT has been further expanded toward digital cytometry. Jo et al. [80] recently demonstrated that 3D RI tomograms can be used to computationally emulate conventional flow cytometry signals. By applying a light-scattering model to the 3D RI maps, they successfully reconstructed forward scattering (FSC) and side scattering (SSC) signals, achieving a seamless digital transition from volumetric imaging to high-throughput statistical analysis without additional hardware

## 3. The AI Revolution: Virtual Staining and Automated Diagnosis

*3.1 3D Virtual Histopathology*

A landmark advancement in 2025 is the application of deep learning to generate 3D virtual Hematoxylin and Eosin (H&E) staining from label-free HT images of thick tissue [60] (Fig. 3a). Traditional histopathology requires labor-intensive sectioning and staining [81], which limits analysis to 2D planes. The new framework utilizes a generative adversarial network (GAN) to transform 3D RI tomograms of unstained, 4–50 μm thick cancer tissues into virtual H&E volumes. This "virtual biopsy" capability preserves microanatomical structures such as glands and nuclei with high fidelity (SSIM > 0.75) compared to chemical staining. It allows pathologists to navigate through volumetric tissue data, offering a more comprehensive assessment of tumor microenvironments without the artifacts associated with physical sectioning.

*3.2 Clinical Diagnostics and Classification*

AI-driven HT is showing high sensitivity in clinical diagnostics and cell classification. In the context of central nervous system (CNS) infections, a DenseNet-based deep learning model analyzed the 3D morphology of immune cells in cerebrospinal fluid (CSF) to diagnose the etiology (viral vs. non-viral) and predict patient prognosis with high accuracy (AUROC > 0.89) [68] (Fig. 3b). Similarly, in biliary tract cancer, EfficientNet-b3 models utilized LD characteristics extracted from HT data to classify cancer cells with >98% accuracy, identifying specific lipid metabolic signatures as diagnostic markers [59]. Furthermore, a ResNet based framework successfully classified distinct cell death pathways—apoptosis, necroptosis, necrosis, and ferroptosis—based solely on label-free morphological features, detecting early changes hours before conventional fluorescence markers [34].

*3.3 Organelle-Aware Representation Learning*

Beyond simple classification, self-supervised learning frameworks are being used to extract biologically meaningful features. "Organelle-aware" representation learning models have been trained to segment and analyze mitochondria and lysosomes in live neurons without fluorescent labels [10] (Fig. 3c). By learning from correlative HT and fluorescence data, these models can predict mitochondrial dysfunction and network fragmentation in neurodegenerative disease models. This approach circumvents the need for exogenous labeling in longitudinal studies, enabling the non-invasive monitoring of subcellular health and the quantification of organelle-specific biophysical parameters like dry mass and density.

## 4. Application Domains of Holotomography in 2025

HT has evolved into a broadly applicable platform for biological and biomedical research, extending beyond its origins in quantitative phase imaging. By reconstructing three-dimensional RI distributions, HT enables label-free visualization of cells and tissues while providing intrinsic, quantitative descriptors related to biomolecular content, mass density, and spatial organization [82–85]. These attributes make HT particularly well suited to experimental contexts in which exogenous labeling is impractical, phototoxicity must be minimized, or unbiased, data-driven phenotypic characterization is required.

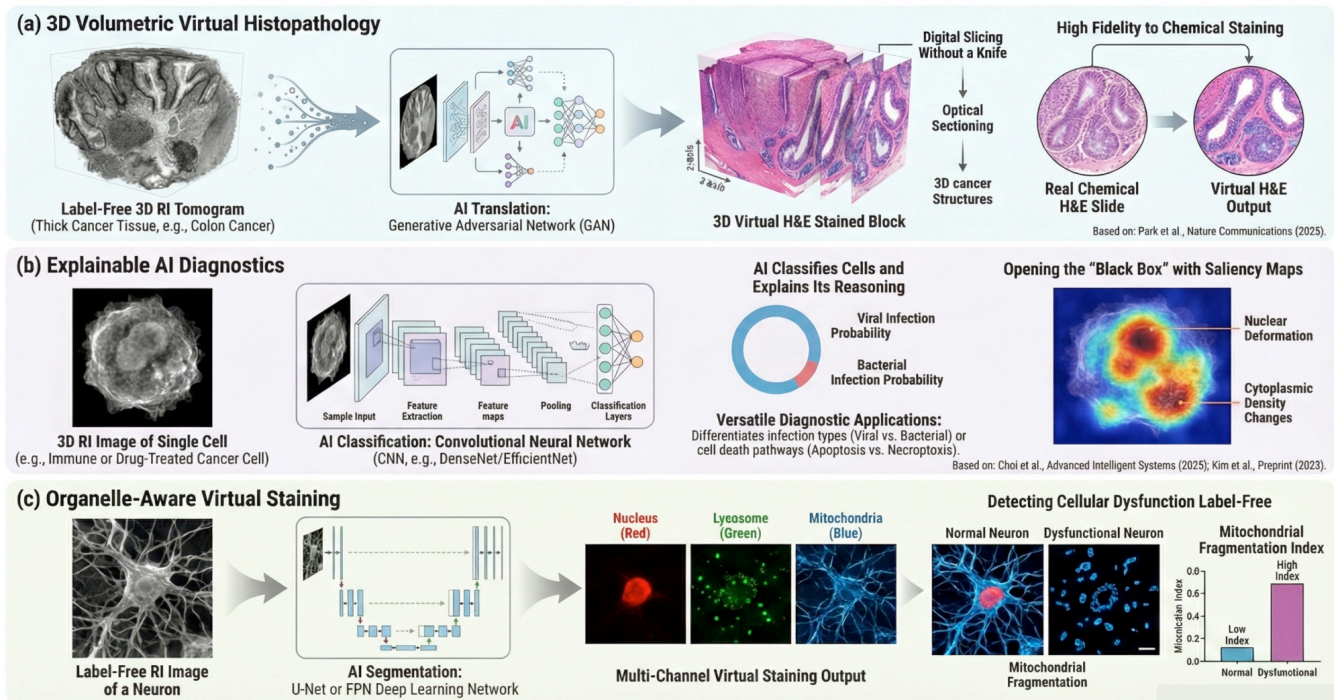

**Figure 3 | AI–enabled HT for virtual histology and automated diagnosis.** (a) Three-dimensional virtual histopathology of thick tissues. By combining HT with generative deep learning, thick cancer tissue sections (up to 50 μm) can be virtually stained in 3D from label-free RI tomograms, producing volumetric H&E-like images that reveal continuous microanatomical features inaccessible to conventional thin-section histology. (b) Explainable AI for clinical diagnostics. Deep learning models applied to 3D RI data enable accurate classification tasks, such as distinguishing viral from non-viral central nervous system infections in cerebrospinal-fluid immune cells. Gradient-based saliency maps indicate that predictions are driven by specific subcellular features, including nuclear morphology and cytoplasmic density. (c) Organelle-aware representation and virtual staining. Self-supervised networks trained on HT data can generate virtual fluorescence channels for subcellular organelles, enabling quantitative analysis of organelle-specific phenotypes, such as mitochondrial fragmentation in live human neurons, without phototoxic labeling. Images are schematic and illustrative; readers are referred to the cited studies for experimental data.

An analysis of studies published in 2025 indicates that HT applications can be grouped into four major biological and clinical domains (Table 1): (A1) cell biology and biophysics, (A2) microbiology and immunology, (A3) cancer and disease biology, and (A4) three-dimensional tissues, organoids, and developmental systems. Although these domains span different biological scales and degrees of translational maturity, they are unified by a common methodological principle: the use of intrinsic physical contrast to characterize biological state, organization, and heterogeneity in three dimensions.

*4.1 Cell Biology and Biophysics*

In 2025, cell biology and biophysics continued to represent a primary application domain of HT, reflecting the technique's ability to quantify intrinsic physical properties of living cells in a label-free and minimally perturbative manner. Through volumetric reconstruction of RI distributions, HT provides direct access to key biophysical parameters, including cell volume, dry mass, spatial mass-density distributions, and subcellular heterogeneity. Together, these descriptors define a cell's physical phenotype and offer quantitative readouts of cellular state, growth, and functional variability that are difficult to obtain using conventional imaging approaches.

Several studies used HT to investigate core principles of cellular homeostasis and regulation. A notable example demonstrated that cells actively regulate cell mass density as a primary control variable [86,87], prioritizing density recovery over volume or total mass during osmotic stress [22]. This work reframed classical views of growth control and highlighted how physical measurements can reveal regulatory strategies that are not apparent from molecular assays alone.

HT has also enabled systematic exploration of subcellular heterogeneity and organelle-level phenotypes. The recent studies reported segmental heterogeneity within mitochondria, linking spatially resolved differences in morphology and density to functional diversity within a single cell [27]. Similarly, lipid droplets were quantitatively characterized in three dimensions, revealing how their abundance and spatial organization change in response to metabolic stress [17]. These observations highlight the utility of HT for studying intracellular organization as a continuous physical landscape rather than a collection of discrete, labeled compartments.

Studies in 2025 linked LD accumulation to fructose-induced adipocyte differentiation [29] and metabolic shifts in bipolar disorder-derived astrocytes treated with lithium [46]. In yeast Yarrowia lipolytica, HT quantified intracellular lipid accumulation to optimize extraction processes [65].

HT has emerged as a powerful tool for quantifying liquid-liquid phase separation (LLPS) in living cells by providing high-resolution RI maps. Lee et al. utilized HT to investigate the condensation of PDIA6, demonstrating that RI values can precisely reflect the protein concentration within these condensates [88]. Their study showcased HT's unique ability to monitor the dynamic formation and dissolution of biomolecular condensates label-free, offering a quantitative approach to understanding insulin folding regulation through phase transition [89–91].

Another important application area involved stress responses and adaptive remodeling. Oxidative stress, metabolic perturbation, and environmental challenges induce subtle yet spatially heterogeneous changes in intracellular composition [92,93]. HT captured these responses through redistribution of RI and mass density, providing a holistic view of how cells reorganize under stress [18]. In particular, stress-induced condensation phenomena associated with Heat Shock Factor 1 were quantitatively analyzed, offering insight into the physical properties of intracellular condensates in living cells [23].

HT has also advanced evolutionary photonics by revealing nature's complex optical designs. Bogdanov et al. used HT to map 3D RI gradients in squid iridocytes [94], deciphering how intracellular modulations enable dynamic light manipulation in their native, hydrated states. Rational nanomaterial design, such as RI–enhanced gold-doped carriers, overcomes intrinsic contrast limitations of soft matter in HT [24]. This strategy establishes HT as a material-aware, multimodal platform for quantitative intracellular tracking.

*4.2 Microbiology and Immunology*

Another major application domain of HT in 2025 lies in microbiology and immunology, where label-free imaging offers distinct advantages. Many microbial systems are difficult to genetically manipulate, and immune cells often undergo rapid, dynamic changes that are challenging to capture with fluorescence-based methods. HT provides a universal, species-agnostic imaging modality capable of addressing these challenges.

In microbiology, HT has been applied to bacterial and fungal pathogens to investigate morphology, growth, and stress adaptation. Three-dimensional imaging revealed how fungal pathogens remodel their cell walls to invade confined microenvironments, directly linking structural changes to invasive growth behavior [37]. In parallel, HT enabled real-time, volumetric monitoring of antimicrobial and antibiofilm activity of newly identified peptides against multidrug-resistant bacteria, revealing progressive morphological disruption and intracellular density changes during bacterial killing [19].

In immunology, HT has been used to characterize immune cell activation and inflammatory responses. Label-free volumetric imaging captured cytoplasmic and nuclear reorganization in macrophages exposed to particulate matter, connecting intracellular structural changes to inflammation-related pathways [32]. Such measurements provide an integrated view of immune activation that complements conventional marker-based immunophenotyping.

HT is effectively used to visualize bacterial cell wall remodeling in *Fusarium oxysporum* [37] and the antibiofilm activity of novel peptides like Hirunipin against multidrug-resistant *Acinetobacter baumannii* [19]. In environmental toxicology, HT assessed the impact of heavy metals on the diatom *Skeletonema pseudocostatum*, utilizing chloroplast RI and lipid content as bioindicators of pollution [62]. In reproductive biology, HT provided a comprehensive quality assessment of cryopreserved stallion spermatozoa, correlating morphological damage with oxidative stress markers [20], and evaluated sperm selected via surface-engineered microfluidic devices [95].

Importantly, HT has also been applied to host–pathogen interaction studies, where simultaneous imaging of host cells and pathogens is required. Because HT does not rely on species-specific labels, it enables unbiased visualization of interacting biological entities within the same imaging volume. These capabilities have been increasingly leveraged in complex infection models, including organoid-based systems discussed further below [21]. Collectively, these applications highlight HT's value as a common physical framework across microbiology and immunology.

*4.3 Cancer and Disease Biology*

In cancer and disease biology, HT has emerged as a quantitative phenotyping tool capable of capturing subtle, label-free signatures associated with pathological states. Cancer cells often exhibit altered nuclear architecture, intracellular organization, and metabolic profiles, all of which are reflected in RI distributions.

In 2025, HT was widely applied to investigate tumor cell heterogeneity. Studies demonstrated that LD content and spatial organization, quantified through three-dimensional imaging, serve as discriminative features in cancer cells subjected to metabolic stress. These findings suggest that HT-derived physical phenotypes may function as biomarkers orthogonal to genomic or proteomic signatures.

In the context of cancer biology, HT provides a dual advantage by capturing both dynamic processes and metabolic states. Cangkrama et al. successfully tracked the label-free mitochondrial transfer between cancer cells and fibroblasts [58], elucidating its role in tumor microenvironment remodeling. Complementing this structural insight, Chen et al. leveraged HT to quantify lipid metabolism markers in pancreatic cancer [49], highlighting its clinical value in identifying metabolic vulnerabilities. Together, these studies position HT as a comprehensive platform for both fundamental cancer research and precision diagnostics. Time-resolved HT uncovered vesicle-mediated mitochondrial clearance as a dynamic pathway complementary to mitophagy in cancer cells [24]. These findings demonstrate HT's power to interrogate systems-level organelle remodeling linked to metabolic vulnerability in disease.

HT has also been used to evaluate therapeutic response and cell fate decisions. Label-free classification frameworks distinguished multiple cell death pathways—including apoptosis, necroptosis, and ferroptosis—based solely on morphological and biophysical features extracted from HT data, detecting early divergence before conventional molecular markers became evident [34]. Such approaches highlight the sensitivity of physical phenotypes to therapeutic perturbations.

Beyond oncology, HT has been applied to non-cancer disease models, particularly in neurology and metabolism. Organelle-aware analyses enabled label-free detection of mitochondrial dysfunction in live human neurons, providing quantitative indicators of early pathological changes without the phototoxicity associated with fluorescent dyes [10]. These studies underscore the relevance of HT in disease biology, where early, non-invasive phenotypic characterization is essential.

*4.4 Three-Dimensional Tissues, Organoids, and Development*

One of the most significant expansions of HT in 2025 has been its application to three-dimensional tissues, organoids, and developmental systems. Improvements in imaging of thick, scattering samples enabled HT to move beyond isolated cells toward more physiologically relevant models.

In organoid research, HT has been used for longitudinal, label-free monitoring of growth and architecture. Intestinal organoids were imaged over extended periods to quantify volumetric growth and protein density changes, providing insights into developmental dynamics and drug response without phototoxicity [44]. Similarly, patient-derived cancer organoids were shown to recapitulate key structural and metabolic features of native tumors, validating their relevance as disease models [8].

HT has also made important contributions to developmental and reproductive biology. Quantitative holographic analysis of cryopreserved spermatozoa revealed correlations between morphological damage, dry mass alterations, and oxidative stress, enabling comprehensive, label-free assessment of reproductive cell quality [20]. In stem cell–derived tissue models, HT tracking enabled identification of endothelial cell identity and zonation based purely on biophysical signatures [16].

HT's label-free volumetric imaging is ideal for monitoring 3D developmental processes. Leong et al. used HT to investigate luminogenesis in mammalian follicles, quantifying lumen expansion and cellular rearrangements via spatiotemporal RI distributions [96]. This highlights HT as a powerful platform for organogenesis, providing high-contrast visualization of internal cavities without invasive labeling.

HT analysis of intestinal stem cells on xenogeneic-free polymer surfaces revealed enhanced migration and cytoskeletal reorganization [32]. In tissue engineering, HT helped evaluate oxygen-releasing hydrogels that promote mitochondrial biogenesis and skin flap regeneration [7]. Additionally, HT was used to characterize liver sinusoidal endothelial cells derived from stem cells [3] and to assess cellular behaviors in volumetrically bioprinted constructs [97].

Perhaps most strikingly, HT has been applied to zoonotic virus infection models in bat organoids. Label-free volumetric imaging enabled visualization of infection dynamics and cytopathic effects across multiple virus species, revealing species-specific susceptibility and tissue responses in a physiologically relevant context [21]. These studies demonstrate the potential of HT to operate at the intersection of developmental biology, infectious disease, and translational research.

## 5. Roadmap and Future Directions

The literature of 2025 highlights that HT is maturing from a specialized research technique into a versatile platform for biomedical discovery and clinical diagnostics. Based on current trends, we propose the following roadmap for the continued evolution of HT (Fig. 4).

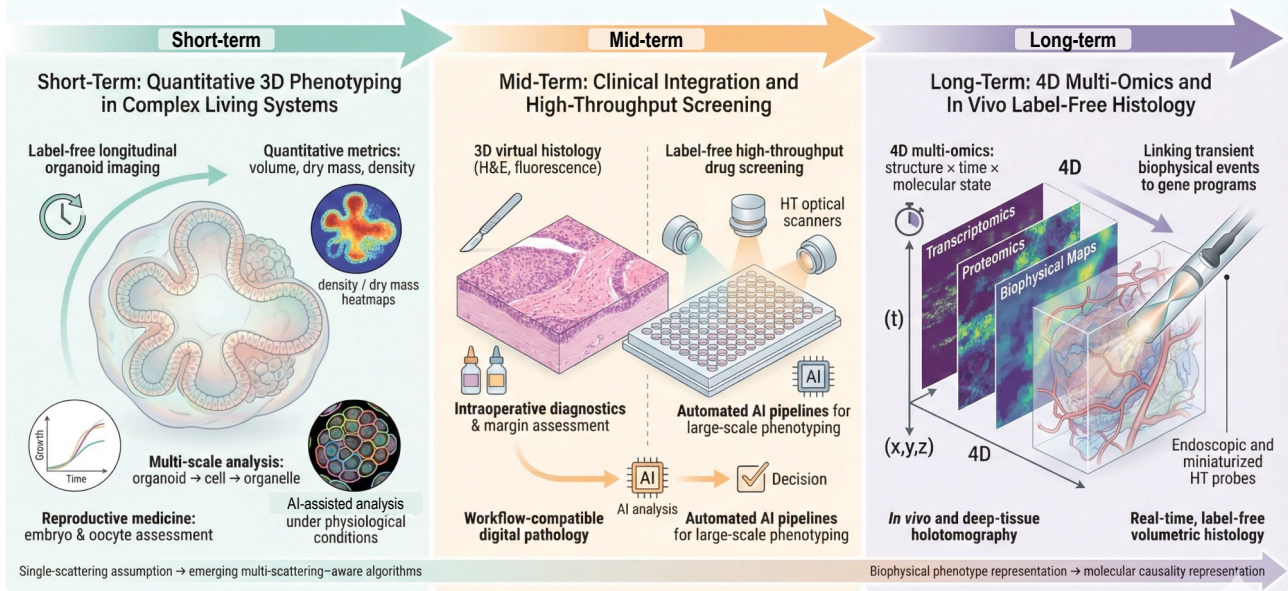

Figure 4 | Strategic roadmap for HT: from biological insight to clinical precision. (a) Short term: AI robustness and standardization. To address domain shifts across cell types and experimental conditions, near-term efforts will focus on RI calibration standards and large-scale foundation models trained on multi-center datasets, enabling robust, organelle-aware segmentation and phenotype classification across diverse contexts. (b) Mid term: Integration into clinical and screening workflows. HT-based 3D virtual H&E staining of thick tissue biopsies may support rapid intraoperative assessment without physical sectioning artifacts. In parallel, automated high-throughput HT platforms are expected to expand in drug screening, using quantitative metrics such as cell mass density and organoid volume to assess therapeutic responses. (c) Long term: 4D multi-omics and in vivo imaging. Future directions include integrating time-resolved 3D RI imaging with spatial transcriptomics and proteomics to generate dynamic multi-omics maps linking biophysical states to gene-expression programs, as well as miniaturized optical implementations enabling *in vivo*, label-free histology.

[Short-Term (1–2 years)]

**Organoid systems:** quantitative, longitudinal, and multi-scale phenotyping. In the near term, organoid research is expected to accelerate markedly, driven by both regulatory shifts toward reduced animal experimentation [98] and rapid advances in organoid culture technologies [99,100] that enable higher reproducibility, scalability, and biological relevance. Within this context, HT offers a unique advantage as a label-free, high-resolution, and quantitative imaging modality, well suited for long-term, time-lapse monitoring of living organoids without phototoxicity or exogenous labeling [101–103]. However, to enable the full utilization of HT in organoid applications, the penetration depth of HT should be further enhanced. Most HT reconstruction rely on the assumption of single scattering, whereas physiologically relevant organoid becomes the size of a few hundred micrometers. Developments of HT algorithms considering multiple light scattering and aberration may expedite and expand the application of HT for the study of organoids [104–106].

Beyond morphological visualization, the quantitative nature of HT is likely to enable systematic phenotyping of organoid growth, maturation, and heterogeneity, including measurements of overall organoid size, volume, and dry mass. Importantly, these capabilities are expected to extend toward multi-scale analysis, capturing not only organoid-level dynamics but also the properties of constituent cells and intracellular organelles, such as nuclear morphology, cytoplasmic density distributions, and subcellular mass redistribution during differentiation or stress responses. As a result, HT-based organoid imaging is poised to become a foundational tool for functional assessment, drug-response profiling, and quality control in next-generation organoid platforms.

**Reproductive medicine:** label-free imaging of oocytes and embryos. Another short-term growth area for HT lies in reproductive medicine, particularly in the imaging of sperms, oocytes and preimplantation embryos [107–112]. Current clinical workflows rely heavily on bright-field morphology and subjective grading, while fluorescence-based methods are largely excluded due to concerns over invasiveness and phototoxicity. HT provides a compelling alternative by enabling non-invasive, label-free, and quantitative 3D imaging of oocytes and embryos under physiologically relevant conditions.

In the near future, HT is expected to facilitate time-lapse monitoring of embryo development, capturing subtle volumetric, mass-density, and intracellular structural changes that are not accessible with conventional imaging. Quantitative features derived from RI distributions—such as cytoplasmic organization, organelle dynamics, and compaction-related density changes—may provide objective biomarkers of developmental competence. When combined with emerging AI-based analysis frameworks, HT-based embryo and oocyte imaging has the potential to support more standardized and data-driven assessment in assisted reproductive technologies, while preserving the non-invasive requirements essential for clinical translation.

[Mid-Term (3–5 years)]

**Clinical integration**: virtual histology and workflow-driven deployment. In the mid-term, the demonstrated capability of 3D virtual staining of H&E and fluorescence labeling using HT [13,60] is expected to drive gradual integration into clinical pathology workflows, particularly in settings where conventional frozen-section preparation and chemical staining introduce artifacts or delays. Potential applications include rapid intraoperative assessment, such as tumor margin evaluation and Mohs surgery, where volumetric, stain-free reconstruction of tissue architecture could complement or, in selected contexts, augment existing diagnostic modalities [113]. Rather than replacing established pathology practices outright, early clinical adoption is likely to focus on adjunctive use cases that benefit from HT's label-free, three-dimensional, and quantitative imaging capabilities [114–116].

Successful clinical translation in this phase will depend on workflow compatibility, including acquisition speed, robustness to tissue heterogeneity, and seamless integration with digital pathology infrastructure. Demonstrations of diagnostic concordance with conventional histology, along with prospective validation in clinically relevant cohorts, will be essential to establish confidence in HT-based virtual staining approaches.

**High-throughput screening: quantitative, label-free metrics at scale.** Concurrently, HT is poised to expand its role in high-throughput drug discovery and toxicology screening, particularly through its ability to quantify cellular and organoid-level mass-density metrics without exogenous labels. Measurements of dry mass, volumetric growth, and density redistribution provide physically grounded readouts of cellular state that complement biochemical assays [117–119]. In organoid-based screening platforms, HT enables longitudinal monitoring of growth, viability, and structural integrity, offering a scalable alternative to endpoint fluorescence assays that are prone to perturbation and phototoxicity.

As screening formats scale to thousands of conditions, automation of acquisition and analysis pipelines will become indispensable. AI-based approaches, including organelle-aware segmentation and phenotyping models, are expected to play a central role in extracting interpretable features from the large volumetric datasets generated in high-throughput settings.

**Standardization and AI generalizability: from proof-of-concept to reproducible deployment.** A key mid-term challenge for both clinical and screening applications is model generalizability under domain shift. Although current deep learning models demonstrate high performance for tasks such as cell death classification and organelle segmentation on internal datasets [120–124], their robustness across different cell types, sample preparation protocols, imaging conditions, and instruments remains largely unexplored. Addressing this limitation will require the development of large-scale, multi-center datasets and coordinated benchmarking efforts to support the training of more robust, transferable foundation models for label-free cellular phenotyping.

In parallel, standardization of RI measurements will be critical for reproducibility across laboratories. Establishing universal calibration protocols, reference phantoms, and consensus criteria for organelle-level RI and texture signatures—such as distinguishing mitochondria from lysosomes—will enable more consistent biological interpretation and facilitate broader adoption of HT-based quantitative imaging.

[Long-Term (5–10 years)]

**4D multi-omics: linking dynamic biophysics to molecular programs.** Over the long term, a major frontier for HT lies in its systematic integration with spatial omics technologies, enabling direct coupling between dynamic biophysical phenotypes and molecular states. Initial studies have begun to correlate HT-derived morphological features with transcriptomic profiles in complex biological systems, including cancer tissues and organoid models. Building on these early efforts, a long-term objective is the development of 4D multi-omics platforms, in which time-resolved, label-free structural information from live-cell HT is temporally aligned with spatial transcriptomic and proteomic measurements.

Such integration would enable researchers to associate transient biophysical events—for example, changes in cytoplasmic density, nuclear compaction, or mitochondrial fragmentation—with downstream gene expression programs and signaling pathways. By providing a physical context for molecular readouts, 4D multi-omics frameworks have the potential to move beyond static correlations and toward a more mechanistic understanding of how cellular structure, mass distribution, and function evolve over time.

*In vivo* **and deep-tissue imaging: extending label-free histology beyond *ex vivo* systems.** A parallel long-term direction involves extending HT from *ex vivo* samples and organoid-scale systems toward *in vivo* and deep-tissue imaging. Achieving this goal will require substantial advances in optical hardware, computational reconstruction, and probe design to address challenges such as multiple scattering, limited numerical aperture, and motion artifacts in living tissues. Progress along these fronts could enable HT imaging at depths beyond current organoid scales or support the development of miniaturized or endoscopic HT probes suitable for clinical use [125,126].

If realized, such capabilities would position HT as a platform for label-free, volumetric histology directly within the patient, complementing conventional biopsy-based diagnostics. While significant technical and regulatory hurdles remain, long-term advances in *in vivo* HT could fundamentally reshape how tissue structure and pathology are assessed, bridging the gap between optical imaging, digital pathology, and real-time clinical decision-making.

The body of work published in 2025 indicates that RI–based tomography has evolved into a broadly applicable framework for physical phenotyping. Through the convergence of physics-informed reconstruction, multimodal optical integration, and AI-enabled semantic analysis, HT enables the characterization of biological state across multiple length scales, from subcellular organization to tissue-level architecture, without the need for exogenous labels. Looking ahead, progress in the field is likely to be driven less by incremental improvements in spatial resolution and more by efforts to translate these quantitative physical descriptors into standardized, interpretable, and ultimately clinically relevant metrics.

## Acknowledgments

We thank all the members of KAIST Biomedical Optics Laboratory and Tomocube Inc. for fruitful discussions. Images were created with BioRender.com, Power Point, and generative AI tools.

## Conflicts of interest

Y.KP. have financial interests in Tomocube Inc., a company that commercializes HT system.

## Funding

This work was supported by the National Basic Science Research Program through the National Research Foundation of Korea (NRF) funded by the Ministry of Science and ICT (RS-2024-00442348), Korea Institute for Advancement of Technology (KIAT) through the International Cooperative R&D program (P0028463).

## References

1. Y. Park, C. Depeursinge, and G. Popescu, "Quantitative phase imaging in biomedicine," Nature Photon **12**, 578–589 (2018).
2. G. Kim, H. Hugonnet, K. Kim, J.-H. Lee, S. S. Lee, J. Ha, C. Lee, H. Park, K.-J. Yoon, Y. Shin, G. Csucs, I. Hitchcock, L. Mackinder, J. H. Kim, T. H. Hwang, S. Lee, P. O'Toole, B.-K. Koo, J. Guck, and Y. Park, "Holotomography," Nat Rev Methods Primers **4**, 51 (2024).
3. L. Bianchi, A. Bresci, K. J. Kobayashi-Kirschvink, G. Paroni, P. Saccomandi, P. T. C. So, and J. W. Kang, "A Multimodal Imaging Framework to Advance Phenotyping of Living Label-free Breast Cancer Cells," JoVE 68498 (2025).
4. J. Oh, H. Hugonnet, W. S. Park, and Y. Park, "Generalized reciprocal diffractive imaging for reference-free, single-shot quantitative phase microscopy," Commun Phys **8**, 383 (2025).
5. J. Hur, K. Kim, S. Lee, H. Park, and Y. Park, "Melittin-induced alterations in morphology and deformability of human red blood cells using quantitative phase imaging techniques," Sci Rep **7**, 9306 (2017).
6. S. Lee, K. Kim, A. Mubarok, A. Panduwirawan, K. Lee, S. Lee, H. Park, and Y. Park, "High-Resolution 3-D Refractive Index Tomography and 2-D Synthetic Aperture Imaging of Live Phytoplankton," J. Opt. Soc. Korea, JOSK **18**, 691–697 (2014).
7. M.-T. Hong, G. Lee, and Y.-T. Chang, "A Non-Invasive, Label-Free Method for Examining Tardigrade Anatomy Using Holotomography," Tomography **11**, 34 (2025).
8. K. Han, J. Choi, C. Kim, S. Kang, H. An, C.-G. Pack, J.-H. Ahn, H. Kwon, C. W. Kim, J. S. Song, T. W. Kim, E. Tak, and J. E. Kim, "Gelatin-Based Soft-Tissue Sarcoma Organoids Recapitulate Patient Tumor Characteristics," Biomater Res **8**, 0293 (2025).
9. J. Park, B. Bai, D. Ryu, T. Liu, C. Lee, Y. Luo, M. J. Lee, L. Huang, J. Shin, Y. Zhang, D. Ryu, Y. Li, G. Kim, H. Min, A. Ozcan, and Y. Park, "Artificial intelligence-enabled quantitative phase imaging methods for life sciences," Nat Methods **20**, 1645–1660 (2023).
10. S. Yoo, E. Yang, S.-H. Kim, H. Park, D. Kim, and M. L. Choi, "Organelle-Aware Representation Learning Enables Label-Free Detection of Mitochondrial Dysfunction in Live Human Neurons," (n.d.).
11. M. Amirola-Martinez, T. Combriat, K. Ferencevic, I. Wilhelmsen, A. Dalmao-Fernandez, P. A. Olsen, J. Stokowiec, A. Aizenshtadt, and S. Krauss, "Aspects of zone-like identity and holotomographic tracking of human stem cell-derived liver sinusoidal endothelial cells," Front. Cell Dev. Biol. **13**, 1528991 (2025).
12. B. Bai, X. Yang, Y. Li, Y. Zhang, N. Pillar, and A. Ozcan, "Deep learning-enabled virtual histological staining of biological samples," Light Sci Appl **12**, 57 (2023).
13. Y. Jo, H. Cho, W. S. Park, G. Kim, D. Ryu, Y. S. Kim, M. Lee, S. Park, M. J. Lee, H. Joo, H. Jo, S. Lee, S. Lee, H. Min, W. D. Heo, and Y. Park, "Label-free multiplexed microtomography of endogenous subcellular dynamics using generalizable deep learning," Nat Cell Biol **23**, 1329–1337 (2021).

14. M. E. Kandel, Y. R. He, Y. J. Lee, T. H.-Y. Chen, K. M. Sullivan, O. Aydin, M. T. A. Saif, H. Kong, N. Sobh, and G. Popescu, "Phase imaging with computational specificity (PICS) for measuring dry mass changes in sub-cellular compartments," Nature communications **11**, 6256 (2020).
15. Y. N. Nygate, M. Levi, S. K. Mirsky, N. A. Turko, M. Rubin, I. Barnea, G. Dardikman-Yoffe, M. Haifler, A. Shalev, and N. T. Shaked, "Holographic virtual staining of individual biological cells," Proceedings of the National Academy of Sciences **117**, 9223–9231 (2020).
16. J. Hong, H. Hugonnet, C. M. Oh, W. S. Park, C. Lee, C. Lee, Y. W. Kim, W. Heo, S.-M. Hong, and Y. Park, "High-speed holotomography of live cells and tissues using multi-pattern sparse axial scanning," Opt. Express **33**, 45708 (2025).
17. H. Khadem, M. Mangini, M. A. Ferrara, A. C. De Luca, and G. Coppola, "Polarization-Sensitive Holotomography for Multidimensional Label-Free Imaging and Characterization of Lipid Droplets in Cancer Cells," Advanced Science **12**, e09420 (2025).
18. M. Machida, S. Kajimoto, R. Shibuya, M. Isono, M. Watabe, Y. Oma, K. Hibino, K. Fujii, M. Okumura, M. Harata, A. Shibata, and T. Nakabayashi, "Lipids Are Involved in Heterochromatin Condensation: A Quantitative Raman and Brillouin Microscopy Study," (2025).
19. S. D. Kumar, J. Park, N. K. Radhakrishnan, Y. P. Aryal, G. Jeong, I. Pyo, B. Ganbaatar, C. W. Lee, S. Yang, Y. Shin, S. Subramaniyam, Y. Lim, S. Kim, S. Lee, S. Y. Shin, and S. Cho, "Novel Leech Antimicrobial Peptides, Hirunipins: Real-Time 3D Monitoring of Antimicrobial and Antibiofilm Mechanisms Using Optical Diffraction Tomography," Advanced Science **12**, 2409803 (2025).
20. M. A. Ferrara, G. Preziosi, R. Boni, R. Ruggiero, and S. C. Gualandi, "Quantitative holographic analysis in stallion spermatozoa following cryopreservation," Sci Rep **15**, 43190 (2025).
21. H. Kim, S.-Y. Heo, Y.-I. Kim, D. Park, Monford Paul Abishek N, S. Hwang, Y. Lee, H. Jang, J.-W. Ahn, J. Ha, S. Park, H. Y. Ji, S. Kim, I. Choi, W. Kwon, J. Kim, K. Kim, J. Gil, B. Jeong, J. C. D. Lazarte, R. Rollon, J. H. Choi, E. H. Kim, S.-G. Jang, H. K. Kim, B.-Y. Jeon, G. Kayali, R. J. Webby, B.-K. Koo, and Y. K. Choi, "Diverse bat organoids provide pathophysiological models for zoonotic viruses," Science **388**, 756–762 (2025).
22. J. Fu, Q. Ni, Y. Wu, A. Gupta, Z. Ge, H. Yang, Y. Afrida, I. Barman, and S. X. Sun, "Cells prioritize the regulation of cell mass density," Science AdvAnceS (2025).
23. S. Kawagoe, M. Matsusaki, T. Mabuchi, Y. Ogasawara, K. Watanabe, K. Ishimori, and T. Saio, "Mechanistic Insights Into Oxidative Response of Heat Shock Factor 1 Condensates," JACS Au **5**, 606–617 (2025).
24. Y. Honda, D. Tokura, S. El Muttaqien, K. Konarita, Y. Kawashima, N. Nishiyama, and T. Nomoto, "Phenylboronic acid-based polymers exerting intracellular hydrophilic-to-hydrophobic conversion to retain within target cells," Chemical Engineering Journal **527**, 171908 (2026).
25. Y. Li, Y. Zhang, M. Chen, S. Antoku, J. Ding, K. Huang, G. G. Gundersen, and W. Chang, "Elevated SUN1 promotes migratory cell polarity defects through mechanically coupling microtubules to the nuclear lamina," Commun Biol (2025).
26. K. Yu, S. T. Chua, A. Smith, A. G. Smith, T. Ellis, and S. Vignolini, "Cultivating Future Materials: Artificial Symbiosis for Bulk Production of Bacterial Cellulose Composites," 2025.04.23.650277 (2025).
27. A. Tsukamura, H. Ariyama, N. Hayashi, S. Miyatake, S. Okado, S. Sultana, I. Terakado, T. Yamamoto, S. Yamanaka, S. Fujii, H. Hamanoue, R. Asano, T. Mizushima, N. Matsumoto, Y. Maruo, and M. Mori, "KNTC1 introduces segmental heterogeneity to mitochondria," Disease Models & Mechanisms **18**, DMM052063 (2025).
28. Y. Lee, W. H. Jung, K. Jeon, E. B. Choi, T. Ryu, C. Lee, D.-N. Kim, and D. J. Ahn, "Membrane-targeted DNA frameworks with biodegradability recover cellular function and morphology from frozen cells," Trends in Biotechnology **43**, 3196–3216 (2025).
29. P. Anantha, X. Wu, S. Elsaid, P. Raj, I. Barman, and S. S. Tee, "Sweet science: Exploring the impact of fructose and glucose on brown adipocyte differentiation using optical diffraction tomography," (2024).
30. P. Rawat, T. Quaderer, I. Karemaker, S. S. Lee, F. Ulliana, Z. Kontarakis, J. E. Corn, and M. Peter, "1 Disruption of nucleolar integrity triggers cellular 2 quiescence through organelle rewiring and secretion," (n.d.).
31. S. Kroschwald, F. Uliana, C. Wilson-Zbinden, A. Timofiiva, J. Zhou, S. S. Lee, L. Gillet, M. Zanella, A. Othman, R. Mezzenga, and M. Peter, "PKA regulates stress granule maturation to allow timely recovery after prolonged starvation," 2025.07.06.663161 (2025).
32. J. Park, J. Lee, H.-J. Kim, S.-H. Chae, J. Shin, J.-H. Lee, Y. Cheon, Y. Jung, S.-K. Mun, J.-J. Kim, S.-H. Kim, G.-S. Hwang, and S. Lee, "Activation of lands cycle-mediated inflammation in living macrophages exposed to label-free particulate matter," Journal of Hazardous Materials **499**, 140027 (2025).
33. P. Anantha, A. Gupta, J. H. Kim, E. Saracino, P. Raj, I. Lucarini, S. Tanwar, J. Chen, L. Gu, J. Agrawal, A. Convertino, and I. Barman, "Disordered Glass Nanowire Substrates Produce in Vivo-Like Astrocyte Morphology Revealed by Low-Coherence Holotomography," Advanced Science e13424 (2025).
34. M. Kim, W. S. Park, G. Kim, S. Oh, J. Do, J. Park, and Y. Park, "Label-free classification of cell death pathways via holotomography-based deep learning framework," (n.d.).
35. Y. Oyama, M. Ohama, N. Yamada, N. Suzuki, K. Kimura, and Y. Hara, "Tetraploid Caenorhabditis elegans embryos exhibit enhanced tolerance to osmotic stress," 2025.10.13.678412 (2025).
36. A. Lauriola, J. H. Enriqué Steinberg, M. Sarubo, E. Maspero, F. A. Rossi, Y. Mouri, M. Pedretti, M. Hajisadeghian, V. Taibi, A. Vettori, N. Vitulo, M. Assfalg, M. D'Onofrio, M. Rossi, A. Yasue, A. Astegno, S. Polo, S. Santi, Y. Kudo, and D. Guardavaccaro, "The E3 ligase RNF32 controls the IκB kinase complex and NF-κB signaling in intestinal stem cells," Molecular Cell **85**, 4254-4267.e9 (2025).
37. H. Miki, M. M. Gomez, A. Itani, D. Yamanaka, Y. Sato, A. Di Pietro, and N. Takeshita, "Cell wall remodeling in a fungal pathogen is required for hyphal growth into microspaces," mBio **16**, e01184-25 (2025).
38. J. Simińska-Stanny, P. Tournier, A. Shavandi, and S. J. Habib, "Geometrical Designs in Volumetric Bioprinting to Study Cellular Behaviors in Engineered Constructs," Adv Healthcare Materials e03550 (2025).
39. E. Ouni, A. Peaucelle, R. Ghasemi, F. Facchinetti, M. Opitz, L. Bigot, A. Sauvat, O. Kepp, F. Jaulin, Y. Loriot, and K. Schauer, "Mechanosensitive interactions of tumoroids with an engineered environment promote cell proliferation and enhance drug response detection," Cell Biomaterials **1**, 100149 (2025).

40. A. E. Melik-Pashaev, D. K. Matveeva, S. V. Buravkov, D. A. Atyakshin, E. S. Kochetova, and E. R. Andreeva, "Microscopy and Image Analysis of Cell-Derived Decellularized Extracellular Matrix," Cell Tiss. Biol. **19**, 33–47 (2025).
41. S. Jeon, S.-H. Jeong, M. H. Lee, J. W. Seo, D.-S. Kim, N. J. Bassous, J. A. Lozano Soto, C. Choi, M. L. Gonzalez, E. B. Nolasco Díaz, H. Kim, S. R. Shin, and J.-U. Park, "Sustained oxygen-releasing hydrogel implants enhance flap regeneration by promoting mitochondrial biogenesis under mild hypoxia," Bioactive Materials **51**, 559–574 (2025).
42. S. Park, S. Y. Sun, J. G. Son, S. Y. Lee, H. K. Shon, O. Kwon, M. Lee, K. Choi, J. Yeun, S. H. Yoon, M. Kim, M. Son, T. G. Lee, and S. G. Im, "Tailored Xenogeneic-Free Polymer Surface Promotes Dynamic Migration of Intestinal Stem Cells," Advanced Materials e13371 (2025).
43. C. Oh, J. Cho, J. Park, H. Lee, and Y. Park, "Morphology-Preserving Holotomography: Quantitative Analysis of 3D Organoid Dynamics," (2025).
44. J. Cho, M. J. Lee, J. Park, J. Lee, S. Lee, C. Chung, B.-K. Koo, Y. Park, and J. Cho, "Label-free, High-Resolution 3D Imaging and Machine Learning Analysis of Intestinal Organoids via Low-Coherence Holotomography," Journal of Visualized Experiments (JoVE) e68529 (2025).
45. D. H. Min, D. Kim, S. T. Hong, J. Kim, M. J. Kim, S. Kwon, A. Kim, and J.-Y. Lee, "Bafilomycin A1 induces colon cancer cell death through impairment of the endolysosome system dependent on iron," Sci Rep **15**, 5148 (2025).
46. G. H. Baek, D. Kim, G. Son, H. Do, G.-B. Yeon, M. J. Lee, M. Ji, J.-H. Son, M. Ju, I. Ahn, C. S. Kang, H. Lee, S. Choi, J. M. Suh, J. Seo, F. H. Gage, M.-J. Paik, Y. Park, D.-S. Kim, and J. Han, "Differential effects of lithium on metabolic dysfunctions in astrocytes derived from bipolar disorder patients," Mol Psychiatry **30**, 5833–5848 (2025).
47. A. Rudawska, B. Szermer-Olearnik, A. Szczygieł, J. Mierzejewska, K. Węgierek-Ciura, P. Żeliszewska, D. Kozień, M. Chaszczewska-Markowska, Z. Adamczyk, P. Rusiniak, K. Wątor, A. Rapak, Z. Pędzich, and E. Pajtasz-Piasecka, "Functionalized Boron Carbide Nanoparticles as Active Boron Delivery Agents Dedicated to Boron Neutron Capture Therapy," IJN **Volume 20**, 6637–6657 (2025).
48. T.-Y. Ha, Y. Kim, S. M. Lim, Y. Hong, and K.-A. Chang, "GPR40 agonist ameliorates neurodegeneration by regulating mitochondria dysfunction and NLRP3 inflammasome in Alzheimer's disease animal models," Biomedicine & Pharmacotherapy **192**, 118678 (2025).
49. Y. Chen, R. Ballarò, M. Sans, F. I. Thege, M. Zuo, R. Dou, J. Min, M. Yip-Schneider, J. Zhang, R. Wu, E. Irajizad, Y. Makino, K. I. Rajapakshe, H. K. Rudsari, M. W. Hurd, R. A. León-Letelier, H. Katayama, E. Ostrin, J. Vykoukal, J. B. Dennison, K.-A. Do, S. M. Hanash, R. A. Wolff, P. A. Guerrero, M. Kim, C. M. Schmidt, A. Maitra, and J. F. Fahrmann, "Long-chain sulfatide enrichment is an actionable metabolic vulnerability in intraductal papillary mucinous neoplasm (IPMN)-associated pancreatic cancers," Gut **74**, 1638–1652 (2025).
50. A. Zhbanov, Y. S. Lee, M. Son, B. J. Kim, and S. Yang, "Microfluidic Electrochemical Impedance Sensor for Hematological Tests of Blood under Different Osmotic Conditions," Anal. Chem. **97**, 21249–21257 (2025).
51. K.-H. Chang, H.-C. Chen, C.-Y. Chen, S.-P. Tsai, M.-Y. Hsu, P.-Y. Wang, S.-Y. Wu, and C.-L. Su, "Natural lignan justicidin A-induced mitophagy as a targetable niche in bladder cancer," Chemico-Biological Interactions **421**, 111723 (2025).
52. J. Ngoenkam, D. Pejchang, T. Nuamchit, U. Wichai, S. Pongcharoen, T. Laorob, and P. Paensuwan, "Nitro Dihydrocapsaicin Attenuates Hyperosmotic Stress-Induced Inflammation in the Corneal Epithelial Cells via SIRT1/Nrf2/HO-1 Pathway," Experimental Eye Research **261**, 110680 (2025).
53. C. Kim, S. Hong, S. H. Ma, J. Lee, H. So, J. Y. Kim, E. Shin, K. Lee, S. Choi, J. Park, Y. Park, Y.-M. Kim, J. H. Kim, and J. Kim, "Replication stress–induced nuclear hypertrophy alters chromatin topology and impacts cancer cell fitness," Proc. Natl. Acad. Sci. U.S.A. **122**, e2424709122 (2025).
54. L. A. Osminkina, P. A. Tyurin-Kuzmin, M. V. Sumarokova, and A. A. Kudryavtsev, "The Impact of Silicon Nanoparticle Porosity on Their Ability to Sensitize Low-Intensity Medical Ultrasound," Sovrem Tehnol Med **17**, 40 (2025).
55. Z. Liu, C. Chu, Y. Chen, C. Chung, F. Mi, M. Ho, W. Hsu, M. Hsieh, M. Chiang, C. Huang, P. Shueng, C. Yang, C. Lee, and C. Lin, "YAP Expression Confers Therapeutic Vulnerability to Cuproptosis in Breast Cancer Cells by Regulating Copper Homeostasis," Adv Healthcare Materials e02769 (2025).
56. M. Ko, J. Ha, S. Kwon, H. E. Lee, J. Y. Mun, D. Yoon, J. Yoo, H. Cho, M. Lee, Y. Lee, S. Bae, J. Y. Lee, J. Y. Kim, S. H. Shin, M. H. Moon, and H. J. Kwon, "Cryptotanshinone Targets HYOU1 to Rewire ER-Mitochondria Communication and Enhance Autophagy in Atherosclerosis," Research Square (2025).
57. P. L. Wang, N. A. Lester, E. N. Perrault, J. Su, D. Gong, C. Shiau, J. Cao, P. T. T. Nguyen, J. W. Bae, D. Olgun, H. I. Hoffman, A. Lam, J. Huang-Gao, S. Rahaman, J. A. Guo, J. L. Barth, N. Caldwell, P. Divakar, J. W. Reeves, A. Bahrami, S. He, M. Patrick, E. Miller, M. Ganci, G. C. Jaramillo, T. S. Hong, J. Y. Wo, H. Roberts, R. Weissleder, H. Choi, C. F. Castillo, K. Cormier, D. T. Ting, T. Jacks, L. Zheng, M. Hemberg, M. Mino-Kenudson, and W. L. Hwang, "The Pdgfd-Pdgfrb axis orchestrates tumor-nerve crosstalk in pancreatic cancer," 2025.08.26.672505 (2025).
58. M. Cangkrama, H. Liu, X. Wu, J. Yates, J. Whipman, C. G. Gäbelein, M. Matsushita, L. Ferrarese, S. Sander, F. Castro-Giner, S. Asawa, M. K. Sznurkowska, M. Kopf, J. Dengjel, V. Boeva, N. Aceto, J. A. Vorholt, and S. Werner, "MIRO2-mediated mitochondrial transfer from cancer cells induces cancer-associated fibroblast differentiation," Nat Cancer **6**, 1714–1733 (2025).
59. S. W. Park, H. C. Moon, S. J. Hong, A. Choi, S.-L. Lee, D. H. Park, E. Shin, J. H. Jo, D. H. Koh, J. Lee, J.-U. Hou, and K. J. Lee, "Enhancing biliary tract cancer diagnosis using AI-driven 3D optical diffraction tomography," Methods **241**, 196–203 (2025).
60. J. Park, S.-J. Shin, G. Kim, H. Cho, D. Ryu, D. Ahn, J. E. Heo, J. R. Clemenceau, I. Barnfather, and M. Kim, "Revealing 3D microanatomical structures of unlabeled thick cancer tissues using holotomography and virtual H&E staining," Nature communications **16**, 4781 (2025).
61. H.-S. Park, H.-G. Choi, I.-T. Jang, T. A. Pham, Z. Jiang, Y.-J. Son, K. Kim, and H.-J. Kim, "Endogenous hepcidin plays an essential role in *Mycobacterium tuberculosis* Rv1876 antigen-induced antimicrobial activity in macrophages," Emerging Microbes & Infections **14**, 2539192 (2025).
62. M. A. Ferrara, E. Cavalletti, V. Bianco, L. Miccio, G. Coppola, P. Ferraro, and A. Sardo, "Holographic tomography of the diatom Skeletonema pseudocostatum used as a bioindicator of heavy metal-polluted waters," PLoS One **20**, e0322960 (2025).

63. W. Dellisanti*, S. Murthy*, E. Bollati, S. P. Sandberg, and M. Kühl, "Moderate levels of dissolved iron stimulate cellular growth and increase lipid storage in Symbiodinium sp.," (2024).
64. J. Park, D. D. Kang, H. Kim, J. H. Oh, and Y. Park, "Peptide PN5 from *Pinus densiflora* Confers in Vivo Protection Against Multidrug-Resistant Salmonella Typhimurium Through Membrane Disruption," ACS Omega acsomega.5c06012 (2025).
65. S. Hong, Y.-E. Jeon, H. Kim, J. Hong, D. Lee, J. H. Park, J. Y. Jung, S. Cha, P. Lee, and J.-S. Hahn, "Pressing extraction: A novel and environmentally sustainable approach to microbial lipid extraction from Yarrowia lipolytica," Separation and Purification Technology **377**, 134148 (2025).
66. F. Salvà-Serra, P. Nimje, B. Piñeiro-Iglesias, L. A. Alarcón, S. Cardew, E. Inganäs, S. Jensie-Markopoulos, M. Ohlén, H.-S. Sailer, C. Unosson, V. Fernández-Juárez, C. O. Pacherres, M. Kühl, E. R. B. Moore, and N. P. Marathe, "Description of Pseudomonas imrae sp. nov., carrying a novel class C β-lactamase gene variant, isolated from gut samples of Atlantic mackerel (Scomber scombrus)," Front. Microbiol. **16**, 1530878 (2025).
67. P. J. Pietras, M. Chaszczewska-Markowska, D. Ghete, A. Tyczewska, and K. Bąkowska-Żywicka, "Saccharomyces cerevisiae recovery from various mild abiotic stresses: Viability, fitness, and high resolution three-dimensional morphology imaging," Fungal Genetics and Biology **178**, 103975 (2025).
68. B. K. Choi, H. H. Yang, J. H. Kim, J. Hong, K. M. Kim, and Y. R. Park, "Deep-Learning Model for Central Nervous System Infection Diagnosis and Prognosis Using Label-Free 3D Immune-Cell Morphology in the Cerebrospinal Fluid," Advanced Intelligent Systems **7**, 2401145 (2025).
69. S. Shin, K. Kim, T. Kim, J. Yoon, K. Hong, J. Park, and Y. Park, "Optical diffraction tomography using a digital micromirror device for stable measurements of 4D refractive index tomography of cells," in G. Popescu and Y. Park, eds. (2016), p. 971814.
70. K. Kim, K. S. Kim, H. Park, J. C. Ye, and Y. Park, "Real-time visualization of 3-D dynamic microscopic objects using optical diffraction tomography," Opt. Express **21**, 32269 (2013).
71. J. Oh, H. Hugonnet, and Y. Park, "Non-interferometric stand-alone single-shot holographic camera using reciprocal diffractive imaging," Nature communications **14**, 4870 (2023).
72. S. Shin, J. Eun, S. S. Lee, C. Lee, H. Hugonnet, D. K. Yoon, S.-H. Kim, J. Jeong, and Y. Park, "Tomographic measurement of dielectric tensors at optical frequency," Nature Materials **21**, 317–324 (2022).
73. H. Hugonnet, S. Shin, and Y. Park, "Regularization of dielectric tensor tomography," Opt. Express **31**, 3774–3783 (2023).
74. H. Hugonnet, M. Lee, S. Shin, and Y. Park, "Vectorial inverse scattering for dielectric tensor tomography: overcoming challenges of reconstruction of highly scattering birefringent samples," Opt. Express **31**, 29654–29663 (2023).
75. J. Lee, H. Son, S. J. Hong, H. Hugonnet, J. Bang, S. Lee, and Y. Park, "Visualizing 3D Anisotropic Molecular Orientation in Polarization Holographic Optical Elements via Dielectric Tensor Tomography," Advanced Optical Materials **12**, 2302346 (2024).
76. J. Lee, B. G. Chae, H. Kim, M. S. Yoon, H. Hugonnet, and Y. K. Park, "High-precision and low-noise dielectric tensor tomography using a micro-electromechanical system mirror," Opt. Express **32**, 23171–23179 (2024).
77. J. Lee, Y. W. Kim, H. Chang, H. Hugonnet, S.-M. Hong, S. Jeon, and Y. Park, "Incoherent dielectric tensor tomography for quantitative 3D measurement of biaxial anisotropy," (2025).
78. C. Oh, "Extending Rytov Approximation to Vector Waves for Tomography of Anisotropic Materials," Phys. Rev. Lett. **134**, (2025).
79. L.-H. Yeh, I. E. Ivanov, T. Chandler, J. R. Byrum, B. B. Chhun, S.-M. Guo, C. Foltz, E. Hashemi, J. A. Perez-Bermejo, and H. Wang, "Permittivity tensor imaging: modular label-free imaging of 3D dry mass and 3D orientation at high resolution," Nature methods **21**, 1257–1274 (2024).
80. J. Jo, H. Hugonnet, M. J. Lee, and Y. Park, "Digital Cytometry: Extraction of Forward and Side Scattering Signals From Holotomography," Journal of Biophotonics **18**, e202400387 (2025).
81. M. Slaoui and L. Fiette, "Histopathology Procedures: From Tissue Sampling to Histopathological Evaluation," in *Drug Safety Evaluation: Methods and Protocols*, J.-C. Gautier, ed. (Humana Press, 2011), pp. 69–82.
82. S. Y. Lee, H. J. Park, C. Best-Popescu, S. Jang, and Y. K. Park, "The Effects of Ethanol on the Morphological and Biochemical Properties of Individual Human Red Blood Cells," PLoS ONE **10**, e0145327 (2015).
83. H. Hugonnet, Y. W. Kim, M. Lee, S. Shin, R. H. Hruban, S.-M. Hong, and Y. Park, "Multiscale label-free volumetric holographic histopathology of thick-tissue slides with subcellular resolution," Adv. Photon. **3**, 026004 (2021).
84. G. Kim, M. Lee, S. Youn, E. Lee, D. Kwon, J. Shin, S. Lee, Y. S. Lee, and Y. Park, "Measurements of three-dimensional refractive index tomography and membrane deformability of live erythrocytes from Pelophylax nigromaculatus," Sci Rep **8**, 9192 (2018).
85. H. Park, T. Ahn, K. Kim, S. Lee, S. Kook, D. Lee, I. B. Suh, S. Na, and Y. Park, "Three-dimensional refractive index tomograms and deformability of individual human red blood cells from cord blood of newborn infants and maternal blood," J. Biomed. Opt **20**, 111208 (2015).
86. G. Popescu, Y. Park, N. Lue, C. Best-Popescu, L. Deflores, R. R. Dasari, M. S. Feld, and K. Badizadegan, "Optical imaging of cell mass and growth dynamics," American Journal of Physiology-Cell Physiology **295**, C538–C544 (2008).
87. R. Barer, "Determination of dry mass, thickness, solid and water concentration in living cells," Nature **172**, 1097–1098 (1953).
88. Y.-H. Lee, T. Saio, M. Watabe, M. Matsusaki, S. Kanemura, Y. Lin, T. Mannen, T. Kuramochi, Y. Kamada, K. Iuchi, M. Tajiri, K. Suzuki, Y. Li, Y. Heo, K. Ishii, K. Arai, K. Ban, M. Hashimoto, S. Oshita, S. Ninagawa, Y. Hattori, H. Kumeta, A. Takeuchi, S. Kajimoto, H. Abe, E. Mori, T. Muraoka, T. Nakabayashi, S. Akashi, T. Okiyoneda, M. Vendruscolo, K. Inaba, and M. Okumura, "Ca2+-driven PDIA6 biomolecular condensation ensures proinsulin folding," Nat Cell Biol **27**, 1952–1964 (2025).
89. B. Wang, L. Zhang, T. Dai, Z. Qin, H. Lu, L. Zhang, and F. Zhou, "Liquid–liquid phase separation in human health and diseases," Sig Transduct Target Ther **6**, 290 (2021).
90. Y. Hong, K. P. Dao, T. Kim, S. Lee, Y. Shin, Y. Park, and D. S. Hwang, "Label-Free Quantitative Analysis of Coacervates via 3D Phase Imaging," Advanced Optical Materials **9**, 2100697 (2021).

91. T. Kim, J. Yoo, S. Do, D. S. Hwang, Y. Park, and Y. Shin, "RNA-mediated demixing transition of low-density condensates," Nat Commun **14**, 2425 (2023).
92. H. Sies, C. Berndt, and D. P. Jones, "Oxidative Stress," Annual Review of Biochemistry **86**, 715–748 (2017).
93. H. J. Forman and H. Zhang, "Targeting oxidative stress in disease: promise and limitations of antioxidant therapy," Nat Rev Drug Discov **20**, 689–709 (2021).
94. G. Bogdanov, A. A. Strzelecka, N. Kaimal, S. L. Senft, S. Lee, R. T. Hanlon, and A. A. Gorodetsky, "Gradient refractive indices enable squid structural color and inspire multispectral materials," (n.d.).
95. S. R. Ghaemi, D. J. Sharkey, N. O. McPherson, M. N. Alemie, K. Vasilev, and S. A. Robertson, "A surface-engineered microfluidic device for antibody-mediated negative selection of high-quality sperm for assisted reproduction," bioRxiv 2025.09.02.673619 (2025).
96. K. W. Leong, Y. Lou, A. Biswas, J. Y. K. Tan, B. H. Ng, X. Lu, X. P. J. Teo, T. B. Lu, C. Bevilacqua, I. Bonne, R. Prevedel, T. Hiraiwa, and C. J. Chan, "Critical phenomenon underlies de novo luminogenesis during mammalian follicle development," (n.d.).
97. E. Teitge, A. Marzi, Á. Barroso, B. Kemper, and J. Schnekenburger, "Analysis of single cell nanoparticle uptake utilizing tomographic phase imaging and fluorescence microscopy," Quantitative Phase Imaging XI 36 (2025).
98. P.-J. H. Zushin, S. Mukherjee, and J. C. Wu, "FDA Modernization Act 2.0: transitioning beyond animal models with human cells, organoids, and AI/ML-based approaches," J Clin Invest **133**, (2023).
99. Y. Ji and Y. Sun, "Advancements in Organoid Culture Technologies: Current Trends and Innovations," Stem Cells and Development **33**, 631–644 (2024).
100. S. A. Yi, Y. Zhang, C. Rathnam, T. Pongkulapa, and K.-B. Lee, "Bioengineering Approaches for the Advanced Organoid Research," Advanced Materials **33**, 2007949 (2021).
101. M. J. Lee, J. Lee, J. Ha, G. Kim, H.-J. Kim, S. Lee, B.-K. Koo, and Y. Park, "Long-term three-dimensional high-resolution imaging of live unlabeled small intestinal organoids via low-coherence holotomography," Exp Mol Med **56**, 2162–2170 (2024).
102. "Diverse bat organoids provide pathophysiological models for zoonotic viruses | Science," https://www.science.org/doi/full/10.1126/science.adt1438.
103. D. Park, D. Lee, Y. Kim, Y. Park, Y.-J. Lee, J. E. Lee, M.-K. Yeo, M.-W. Kang, Y. Chong, S. J. Han, J. Choi, J.-E. Park, Y. Koh, J. Lee, Y. Park, R. Kim, J. S. Lee, J. Choi, S.-H. Lee, B. Ku, D. H. Kang, and C. Chung, "Cryobiopsy: A Breakthrough Strategy for Clinical Utilization of Lung Cancer Organoids," Cells **12**, 1854 (2023).
104. J. Kim, B. Bolton, K. Moshksayan, R. Khanna, M. E. Swartz, M. Ziemczonok, M. Kamra, K. A. Jorn, S. H. Parekh, M. Kujawińska, J. Eberhart, E. S. Cenik, A. Ben-Yakar, and S. Chowdhury, "Inverse-scattering in biological samples via beam-propagation," 2025.08.17.670744 (2025).
105. C. Oh, H. Hugonnet, M. Lee, and Y. Park, "Digital aberration correction for enhanced thick tissue imaging exploiting aberration matrix and tilt-tilt correlation from the optical memory effect," Nat Commun **16**, 1685 (2025).
106. Y. Chung, H. Hugonnet, S.-M. Hong, and Y. Park, "Fourier space aberration correction for high resolution refractive index imaging using incoherent light," Opt. Express, OE **32**, 18790–18799 (2024).
107. C. Lee, G. Kim, T. Shin, S. Lee, J. Y. Kim, K. H. Choi, J. Do, J. Park, J. Do, J. H. Kim, and Y. Park, "Label-free 3D subcellular phenotyping of mouse embryos by holotomography enables early prediction of blastocyst formation," 2024.05.07.592317 (2025).
108. H. Jiang, J. Kwon, S. Lee, Y.-J. Jo, S. Namgoong, X. Yao, B. Yuan, J. Zhang, Y.-K. Park, and N.-H. Kim, "Reconstruction of bovine spermatozoa substances distribution and morphological differences between Holstein and Korean native cattle using three-dimensional refractive index tomography," Sci Rep **9**, 8774 (2019).
109. M. E. Kandel, M. Rubessa, Y. R. He, S. Schreiber, S. Meyers, L. Matter Naves, M. K. Sermersheim, G. S. Sell, M. J. Szewczyk, N. Sobh, M. B. Wheeler, and G. Popescu, "Reproductive outcomes predicted by phase imaging with computational specificity of spermatozoon ultrastructure," Proceedings of the National Academy of Sciences **117**, 18302–18309 (2020).
110. T. H. Nguyen, M. E. Kandel, M. Rubessa, M. B. Wheeler, and G. Popescu, "Gradient light interference microscopy for 3D imaging of unlabeled specimens," Nat Commun **8**, 210 (2017).
111. G. Dardikman-Yoffe, S. K. Mirsky, I. Barnea, and N. T. Shaked, "High-resolution 4-D acquisition of freely swimming human sperm cells without staining," Science Advances **6**, eaay7619 (2020).
112. N. Goswami, N. Winston, W. Choi, N. Z. E. Lai, R. B. Arcanjo, X. Chen, N. Sobh, R. A. Nowak, M. A. Anastasio, and G. Popescu, "EVATOM: an optical, label-free, machine learning assisted embryo health assessment tool," Commun Biol **7**, 268 (2024).
113. L. Wang, M. Li, and T. H. Hwang, "The 3D Revolution in Cancer Discovery," Cancer Discov **14**, 625–629 (2024).
114. J. Park, S.-J. Shin, J. Shin, A. J. Lee, M. Lee, M. J. Lee, G. Kim, J. E. Heo, K. S. Lee, and Y. Park, "Quantification of structural heterogeneity in H&E stained clear cell renal cell carcinoma using refractive index tomography," Biomed. Opt. Express, BOE **14**, 1071–1081 (2023).
115. A. J. Lee, H. Hugonnet, Y. S. Kim, J.-G. Kim, M. Lee, T. Ku, and Y. Park, "Volumetric Refractive Index Measurement and Quantitative Density Analysis of Mouse Brain Tissue with Sub-Micrometer Spatial Resolution," Advanced Photonics Research **4**, 2300112 (2023).
116. Y. Chung, G. Kim, A.-R. Moon, D. Ryu, H. Hugonnet, M. J. Lee, D. Shin, S.-J. Lee, E.-S. Lee, and Y. Park, "Label-free histological analysis of retrieved thrombi in acute ischemic stroke using optical diffraction tomography and deep learning," Journal of Biophotonics **16**, e202300067 (2023).
117. E. R. Polanco, T. E. Moustafa, A. Butterfield, S. D. Scherer, E. Cortes-Sanchez, T. Bodily, B. T. Spike, B. E. Welm, P. S. Bernard, and T. A. Zangle, "Multiparametric quantitative phase imaging for real-time, single cell, drug screening in breast cancer," Commun Biol **5**, 794 (2022).
118. Y. Liu and S. Uttam, "Perspective on quantitative phase imaging to improve precision cancer medicine," JBO **29**, S22705 (2024).
119. A. Marzi, K. M. Eder, Á. Barroso, B. Kemper, and J. Schnekenburger, "Quantitative Phase Imaging as Sensitive Screening Method for Nanoparticle-Induced Cytotoxicity Assessment," Cells **13**, 697 (2024).
120. M. Kim, W. S. Park, G. Kim, S. Oh, J. Do, J. Park, and Y. Park, "Label-free classification of cell death pathways via holotomography-based deep learning framework," 2025.06.06.658404 (2025).

121. C. Hu, S. He, Y. J. Lee, Y. He, E. M. Kong, H. Li, M. A. Anastasio, and G. Popescu, "Live-dead assay on unlabeled cells using phase imaging with computational specificity," Nature communications **13**, 713 (2022).
122. M. J. Lee, G. Kim, M. S. Lee, J. W. Shin, J. H. Lee, D. H. Ryu, Y. S. Kim, Y. Chung, K. Kim, and Y. Park, "Functional immune state classification of unlabeled live human monocytes using holotomography and machine learning," 2023.09.12.557503 (2025).
123. H. Kim, G. Kim, H. Park, M. J. Lee, Y. Park, and S. Jang, "Integrating holotomography and deep learning for rapid detection of NPM1 mutations in AML," Scientific reports **14**, 23780 (2024).
124. G. Kim, D. Ahn, M. Kang, J. Park, D. Ryu, Y. Jo, J. Song, J. S. Ryu, G. Choi, H. J. Chung, K. Kim, D. R. Chung, I. Y. Yoo, H. J. Huh, H. Min, N. Y. Lee, and Y. Park, "Rapid species identification of pathogenic bacteria from a minute quantity exploiting three-dimensional quantitative phase imaging and artificial neural network," Light Sci Appl **11**, 190 (2022).
125. Z. Guang, S. Bharadwaj, Z. Zhang, S. Neill, J. J. Olson, and F. E. Robles, "Label-Free, Real-Time, In Vivo Optical Biopsy with a Handheld Quantitative Phase Microscope," 2025.09.17.675943 (2025).
126. K. Kim, K. Choe, I. Park, P. Kim, and Y. Park, "Holographic intravital microscopy for 2-D and 3-D imaging intact circulating blood cells in microcapillaries of live mice," Sci Rep **6**, 33084 (2016).
15